\documentclass[amsmath, amssymb, aps, pre, reprint, superscriptaddress]{revtex4-1}

\usepackage{graphicx}
\usepackage{dcolumn}
\usepackage{bm}
\usepackage{braket}
\usepackage{color}
\usepackage[colorlinks=true, linkcolor=magenta, citecolor=blue, urlcolor=magenta]{hyperref}

\begin{document}

\title{Fluctuation theorem in cavity quantum electrodynamics systems}

\author{Tatsuro Yuge}
\email[]{yuge.tatsuro@shizuoka.ac.jp}
\affiliation{Department of Physics, Shizuoka University, Suruga, Shizuoka 422-8529, Japan}

\author{Makoto Yamaguchi}
\affiliation{Department of Physics, Tokai University, Hiratsuka, Kanagawa 259-1292, Japan}

\date{\today}

\begin{abstract}
We derive an integral fluctuation theorem (FT) in a general setup of cavity quantum electrodynamics systems. In the derivation, a key difficulty lies in a diverging behavior of entropy change arising from the zero-temperature limit of an external bath, which is required to describe the cavity loss. We solve this difficulty from the viewpoint of absolute irreversibility and find that two types of absolute irreversibility contribute to the integral FT. Furthermore, we show that, in a stationary and small cavity-loss condition, these contributions have simple relationships to the average number of photons emitted out of the cavity, and the integral FT yields an approximate form independent of the setup details. We illustrate the general results with a numerical simulation in a model of quantum heat engine.
\end{abstract}

\maketitle

\section{Introduction}

Fluctuation theorem (FT) is one of the key ingredients in modern statistical physics and thermodynamics \cite{Evans_Searles_02, Esposito_etal_09, Campisi_etal_11, Seifert_12, Broeck_Esposito_15}.
It is a standard approach to formulate a detailed FT and then to derive an integral FT and a second-law-like inequality from it.
In the detailed FT, the probability $P$ of a trajectory (state history) $\Gamma$ during a time interval $\tau$ is compared with the probability $\bar{P}$ of a backward trajectory $\bar{\Gamma}$ subject to a time-reversed dynamics \cite{Crooks_00, Seifert_05, Esposito_Broeck_10, Horowitz_12, Horowitz_Parrondo_13}:
\begin{align}
\frac{\bar{P}(\bar{\Gamma})}{P(\Gamma)} = e^{-\Delta \sigma(\Gamma)}.
\label{intro:detailedFT}
\end{align}
It thus quantifies the stochastic reversibility at the trajectory level in terms of the total entropy change $\Delta \sigma$ along $\Gamma$.
When the system is in contact with a heat bath at temperature $T$, $\Delta \sigma(\Gamma)$ includes the entropy change in the bath, $Q(\Gamma) / T$.
Here, $Q$ denotes the heat transfer from the system to the bath.

Suppose that the bath is in a zero-temperature state ($T \to +0$).
Then, $Q / T$ diverges and thus fully dominates $\Delta \sigma$.
Since this causes the right-hand side of Eq.~(\ref{intro:detailedFT}) to be either zero or positive infinity,
it becomes difficult to derive the integral FT.
Moreover, since the average of the total entropy change goes to infinity, the second law is automatically valid but rather useless.

These singular behaviors are manifested even when the bath temperature is not zero.
At nonzero but extremely low $T$,
although the ordinary integral FT, $\braket{e^{-\Delta \sigma}} = 1$, is theoretically derived,
it is difficult to demonstrate it in experiments and numerical simulations.
This is because it becomes highly rare to observe trajectories in which heat flows from the bath to the system
while such trajectories have extremely large contribution of $e^{-\Delta \sigma(\Gamma)}$ and are therefore essential for demonstration.
Moreover, the validity of the ordinary second law, $\braket{\Delta \sigma} \ge 0$, derived from the ordinary integral FT is obvious
since $\braket{\Delta \sigma}$ is an extremely large positive value.
Therefore, it is desirable to derive an experimentally and numerically accessible integral FT and a useful second-law-like inequality that are free from the diverging behavior of $Q / T$.

In the present paper, focusing on cavity and circuit quantum electrodynamics (cQED) systems, 
we give a solution to these difficulties.
The cQED systems \cite{Raimond_etal_01, Walther_etal_06, Xiang_etal_13, Cottet_etal_17, Masuyama_etal_18, Naghiloo_etal_18} are a typical one where a zero-temperature bath plays a role.
The cavity loss, an inevitable decay factor in these systems,
is caused by the interaction between the cavity photons and the electromagnetic fields in an external system, which we call photon drain.
Theoretically, the photon drain is usually modeled by a zero-temperature state
in order that it only receives photons out of the cavity but does not inject ones into the cavity.
Such a theoretical description is well justified because the excitation energies of the cQED system are usually much larger than the drain temperature.

We find that the cavity loss induces the absolute irreversibility \cite{Murashita_etal_14, Ashida_etal_14, Funo_etal_15, Murashita_15, Murashita_Ueda_17, Murashita_etal_17, Hoang_etal_16, Monsel_etal_18, Manikandan_etal_19} in the zero-temperature limit, and it is the origin of the difficulties.
A trajectory is said to be absolutely irreversible if it is not reversible even stochastically.
That is, the system has absolute irreversibility if there exists a trajectory satisfying either
\begin{align}
P(\Gamma) = 0~~\text{and}~~\bar{P}(\bar{\Gamma}) > 0
\label{forward_irreversibility}
\end{align}
or
\begin{align}
P(\Gamma) > 0~~\text{and}~~\bar{P}(\bar{\Gamma}) = 0.
\label{backward_irreversibility}
\end{align}
In the original proposal \cite{Murashita_etal_14}, only the former type is referred to as absolute irreversibility,
and the latter may be referred to as ergodic inconsistency (or the lack of ergodic consistency) \cite{Evans_Searles_02}.
In the present paper, we refer to the both types as absolute irreversibility.

By using the method for handling the difficulty due to absolute irreversibility \cite{Murashita_etal_14,Murashita_15}, 
we derive a modified integral FT,
\begin{align}
\bigl\langle e^{-\Delta S} \bigr\rangle = 1 - \lambda + \Lambda,
\label{integral_FT}
\end{align}
and the second-law-like inequality for the entropy change $\Delta S$ without the drain's contribution.
In Eq.~(\ref{integral_FT}), $\lambda$ quantifies the absolute irreversibility of the type in Eq.~(\ref{forward_irreversibility}), and $\Lambda$ quantifies that of the type in Eq.~(\ref{backward_irreversibility}).
In contrast to previous works \cite{Murashita_etal_14, Ashida_etal_14, Funo_etal_15, Murashita_15, Murashita_Ueda_17, Murashita_etal_17, Hoang_etal_16, Monsel_etal_18, Manikandan_etal_19}---where only $\lambda$ appears---we find that both $\lambda$ and $\Lambda$ equally contribute to the modified FT in the setup of the present paper.
This comes from the difference in the origins of absolute irreversibility.
The origin in the present setup lies in asymmetry in the dynamics,
whereas that in the previous works lies in restrictions on the forward trajectories.
We furthermore find that, in a steady-state and small $\kappa$ (cavity loss rate) condition, 
$\lambda$ and $\Lambda$ have simple relationships to the average number of emitted photons during the interval $\tau$,
and their combination in the FT (\ref{integral_FT}) yields $-\lambda + \Lambda \simeq \kappa \tau$.
In this condition, therefore, the modified integral FT has an approximate form independent of the details of the setup.

The cQED systems are a potential testbed for quantum thermodynamics \cite{Horowitz_12, Horowitz_Parrondo_13, Cottet_etal_17, Masuyama_etal_18, Naghiloo_etal_18, Scovil_Schulz-DuBois_59, Boukobza_Tannor_06, Scully_atal_11, Dorfman_etal_13, Goswami_Harbola_13, Gong_eatl_16, Dag_etal_16, Murashita_etal_17, Yuge_etal_17, Li_etal_17, Kosloff_13, Mahler_15, Vinjanampathy_Anders_16, Goold_etal_16, Binder_etal_18}.
The present paper therefore have significance in exploring experimental realization of quantum thermodynamics.
We have to exclude the contribution of the zero-temperature bath to the entropy change 
in order to obtain meaningful results free from the diverging behaviors.
And the modifications appear in the FT and the second-law-like inequality.
The latter may affect the thermodynamic bound on the efficiency of quantum heat engines in the cQED systems.

The present paper is organized as follows.
In Sec.~\ref{sec:setup}, we describe a general setup of a cQED system in contact with a photon drain.
The dynamics of the system is governed by the quantum master equation (QME).
We introduce an additional term related to the cavity loss in the QME to have a counterpart of the local detailed balance condition for the photon drain.
Then we briefly review an expression of the QME time evolution in terms of the quantum jump trajectory (stochastic unravelling) to have the probability of a trajectory.
In Sec.~\ref{sec:detailed_FT}, we derive the detailed FT in our setup in a standard method with a time-reversed dynamics.
Then we confirm the appearance of the two types of absolute irreversibility corresponding to Eqs.~(\ref{forward_irreversibility}) and (\ref{backward_irreversibility}).
Section~\ref{sec:integral_FT} is the main part of the present paper.
We derive the modified integral FT (\ref{integral_FT}) and the second-law-like inequality in the absolutely irreversible situation of the setup.
We also investigate a steady-state and small cavity-loss situation.
In Sec.~\ref{sec:example}, we numerically demonstrate the general results in a model of cQED system.
Section~\ref{sec:conclusion} is devoted to the conclusion.
We set $\hbar = 1$ and $k_{\text{B}} = 1$ throughout the present paper.

\section{Setup}\label{sec:setup}

\begin{figure}[t]
\begin{center}
\includegraphics[width=\linewidth]{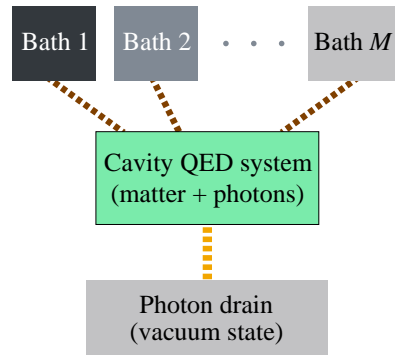}
\end{center}
\caption{Setup of an open cavity QED system.
The system is open due to the interaction with $M$ heat baths and a photon drain.
}
\label{fig:setup}
\end{figure}

We consider a general setup of open cQED systems. We show its schematic diagram in Fig.~\ref{fig:setup}.

The cQED system is composed of matter and cavity photons.
In the general setup, we do not specify the matter part.
For simplicity, we assume the single-mode cavity.

The cQED system interacts with heat baths and a photon drain.
Each of the heat baths is in its own thermal equilibrium state,
and the photon drain is in the vacuum state.
Although we can regard the drain as a zero-temperature heat bath,
we distinguish it from the other baths for clarity of the argument.
We assume that the interaction is weak and the baths and drain are large.
Therefore, the states of the baths and drain do not change in the characteristic time scale of the open cQED system.

\subsection{Quantum master equation}

The dynamics of this open cQED system is governed by the Lindblad-type QME \cite{Breuer_Petruccione_02, Carmichael_93, Wiseman_Milburn_09}:
$\dot{\hat{\rho}} = \mathcal{L} \hat{\rho}$,
where $\hat{\rho}$ is the state (density operator) of the cQED system.
Here, the Liouvillian $\mathcal{L}$ is defined as
\begin{align}
\mathcal{L} \hat{\rho} = -i [\hat{H}, \hat{\rho}] + \sum_{r,k,l}\mathcal{D}[\hat{L}_{r,k,l}] \hat{\rho}
+ \kappa \mathcal{D}[\hat{c}] \hat{\rho},
\label{QME}
\end{align}
where $\mathcal{D}[\hat{A}] \hat{\rho} = \hat{A} \hat{\rho} \hat{A}^\dag - (1/2) (\hat{A}^\dag \hat{A} \hat{\rho} + \hat{\rho} \hat{A}^\dag \hat{A})$.

The first term on the right-hand side of Eq.~(\ref{QME}) describes the unitary time evolution of the cQED system.
$\hat{H}$ is the Hamiltonian of the cQED system.
Its $k$th energy eigenvalue and eigenstate are denoted by $E_k$ and $\ket{E_k}$, respectively.

The second term describes the dissipative time evolution induced by the heat baths.
$\hat{L}_{r,k,l}$ is the jump operator that describes the quantum jump from $\ket{E_k}$ to $\ket{E_l}$ induced by the $r$th bath.
$\hat{L}_{r,k,l}$ satisfies $[\hat{L}_{r,k,l}, \hat{H}] = \omega_{kl} \hat{L}_{r,k,l}$ 
and the local detailed balance (LDB) condition:
\begin{align}
\hat{L}_{r,k,l} = \hat{L}_{r,l,k}^\dag e^{\beta_r \omega_{kl} / 2},
\label{LDB}
\end{align}
where $\omega_{kl} = E_k - E_l$ and $\beta_r$ is the inverse temperature of the $r$th bath.

The third term describes the dissipative time evolution induced by the photon drain.
The prefactor $\kappa$ is the decay constant of the cavity, and $\hat{c}$ is the annihilation operator of the cavity photon.
In other words, this term removes one photon in the cavity with probability $\kappa$ per time,
and it thus represents the cavity-loss effect.
Unlike the second term, the LDB condition is invalid for this term
because the term that describes the opposite jump induced by the photon drain is absent in Eq.~(\ref{QME}).
This results from the assumption that the drain is in the vacuum state (zero-temperature state).

To derive the fluctuation theorem, however, the LDB-like condition is crucial.
We therefore introduce an additional term $\varepsilon \kappa \mathcal{D}[\hat{c}^\dag]$ that describes the opposite jump:
\begin{align}
\mathcal{L}_\varepsilon = \mathcal{L} + \varepsilon \kappa \mathcal{D}[\hat{c}^\dag].
\label{L_e}
\end{align}
We will take the limit of $\varepsilon \to +0$ at the final stage of the calculation.
This limit corresponds to the vacuum (zero-temperature) limit of the drain.
The third term in Eq.(\ref{QME}) plus this additional term reads
$\kappa \mathcal{D}[\hat{c}] + \varepsilon \kappa \mathcal{D}[\hat{c}^\dag]
= \mathcal{D}[\hat{L}_-] + \mathcal{D}[\hat{L}_+]$, where $\hat{L}_- = \sqrt{\kappa} \hat{c}$ and
$\hat{L}_+ = \sqrt{\varepsilon \kappa} \hat{c}^\dag$.
And these jump operators satisfy
\begin{align}
\hat{L}_+ = \hat{L}_-^\dag \varepsilon^{1/2}.
\label{LDB_e}
\end{align}
This is a counterpart of the LDB condition (\ref{LDB}).
From this correspondence, we can introduce the drain temperature $T_{\text{d}}$ and an effective excitation energy $E_{\text{ex}} (> 0)$ to assign $\varepsilon = e^{-E_{\text{ex}} / T_{\text{d}}}$.
We again confirm that the $\varepsilon \to +0$ limit is consistent with the zero-temperature limit $T_{\text{d}} \to +0$.

\subsection{Quantum jump trajectory}

We can express the time evolution of the open system by using the quantum jump trajectory (QJT) \cite{Breuer_Petruccione_02, Carmichael_93, Wiseman_Milburn_09}.
To see this, we note that the state of the system at time $\tau$ can be represented by \cite{Carmichael_93}
\begin{widetext}
\begin{align}
e^{\mathcal{L}_\varepsilon \tau} \hat{\rho}_{\text{ini}}
= \sum_{N=0}^\infty \sum_{j_1, ..., j_N} \int_0^\tau \int_0^{t_N} \dotsi \int_0^{t_2}
\prod_{n=1}^N \Bigl[ \mathcal{U}_{\text{eff}}(\Delta t_n) \mathcal{J}_{j_n} \Bigr] \mathcal{U}_{\text{eff}}(\Delta t_0) \hat{\rho}_{\text{ini}},
\label{time_ev_ens}
\end{align}
where $\Delta t_n = t_{n+1} - t_n$ with $t_{N+1}=\tau$ and $t_0=0$, and $\hat{\rho}_{\text{ini}}$ is the initial state at $t=0$.
The superoperator $\mathcal{U}_{\text{eff}}(\Delta t)$ describes the non-unitary time evolution during an interval with no quantum jump observed:
\begin{align}
\mathcal{U}_{\text{eff}}(\Delta t) \hat{\rho}
&= \hat{U}_{\text{eff}}(\Delta t) \hat{\rho} \hat{U}_{\text{eff}}^\dag(\Delta t),
\\
\hat{U}_{\text{eff}}(\Delta t) &= e^{-i \hat{H}_{\text{eff}} \Delta t},
\\
\hat{H}_{\text{eff}} &= \hat{H} - \frac{i}{2} \sum_{r,k,l} \hat{L}_{r,k,l}^\dag \hat{L}_{r,k,l}
- \frac{i}{2} \sum_{j=\pm} \hat{L}_j^\dag \hat{L}_j.
\end{align}
The superoperator $\mathcal{J}_{j_n}$, on the other hand, describes the state change due to the $n$th quantum jump:
\begin{align}
\mathcal{J}_{j_n} \hat{\rho} = \hat{L}_{j_n} \hat{\rho} \hat{L}_{j_n}^\dag dt_n.
\end{align}
Here, $j_n$ represents the type of the jump. That is, $j_n = (r_n,k_n,l_n)$ if it is induced by the $r_n$th bath and $j_n = \pm$ if it is induced by the drain.

Suppose that, at time $\tau$, we perform an ideal measurement whose measurement basis set is $\{ \ket{\varphi} \}_\varphi$.
Then, writing the initial state in the spectral decomposition, $\hat{\rho}_{\text{ini}} = \sum_\psi p(\psi) \ket{\psi} \bra{\psi}$,
we can express the probability to find the system in one of the measurement basis states $\ket{\varphi}$ as
\begin{align}
\bra{\varphi} e^{\mathcal{L}_\varepsilon \tau} \hat{\rho}_{\text{ini}} \ket{\varphi}
&= \sum_\psi \sum_{N=0}^\infty \sum_{j_1, ..., j_N} \int_0^\tau \int_0^{t_N} \dotsi \int_0^{t_2} P_\varepsilon (\Gamma),
\\
P_\varepsilon (\Gamma)
&= \biggl| \bra{\varphi} \prod_{n=1}^N \Bigl[ \hat{U}_{\text{eff}}(\Delta t_n) \hat{L}_{j_n} \Bigr] \hat{U}_{\text{eff}}(\Delta t_0) \ket{\psi} \biggr|^2 p(\psi) dt^N,
\label{prob_fw_trajectory}
\end{align}
\end{widetext}
where $dt^N = dt_N \cdots dt_1$.
$\Gamma = \{ \psi, (t_1, j_1), ..., (t_N, j_N), \varphi \}$ represents a QJT.
We can specify it with the initial state $\ket{\psi}$ (at $t=0$), the final state $\ket{\varphi}$ (at $t=\tau$), and when and what types of quantum jumps occur during $(0, \tau)$.
From the above equations, we can interpret $P_\varepsilon (\Gamma)$ as the probability density of $\Gamma$.

\section{Detailed Fluctuation Theorem}\label{sec:detailed_FT}

In this section, we derive the detailed FT in our setup,
following the standard approach in the quantum stochastic thermodynamics at the trajectory level \cite{Horowitz_12, Horowitz_Parrondo_13, Murashita_etal_17}.
In this approach, we consider the backward QJT in the time-reversed dynamics.

\subsection{Time-reversed dynamics}

The time-reversal of the QME (\ref{L_e}) is given by
$\dot{\bar{\rho}} = \bar{\mathcal{L}}_\varepsilon \bar{\rho}$, where
\begin{align}
\bar{\mathcal{L}}_\varepsilon \bar{\rho}
= -i [\bar{H}, \bar{\rho}] + \sum_{r,k,l} \mathcal{D}[\bar{L}_{r,k,l}] \bar{\rho}
+ \kappa \mathcal{D}[\hat{c}] \bar{\rho} + \varepsilon \kappa \mathcal{D}[\hat{c}^\dag] \bar{\rho}.
\label{TR-QME}
\end{align}
Here, $\bar{H} = \Theta \hat{H} \Theta^\dag$ and $\bar{L}_{r,k,l} = \Theta \hat{L}_{r,k,l} \Theta^\dag$
with $\Theta$ being the time-reversal operator.

Similarly to the previous section, we can construct a QJT description of the time-reversed evolution
with the time-reversals of the corresponding (super)operators:
\begin{align}
\bar{\mathcal{U}}_{\text{eff}}(\Delta t) \bar{\rho}
&= \bar{U}_{\text{eff}}(\Delta t) \bar{\rho} \bar{U}_{\text{eff}}^\dag(\Delta t),
\notag
\\
\bar{U}_{\text{eff}}(\Delta t) &= \Theta \hat{U}_{\text{eff}}(\Delta t) \Theta^\dag,
\label{TR-U_eff}
\\
\bar{\mathcal{J}}_j \bar{\rho} &= \bar{L}_j \bar{\rho} \bar{L}_j^\dag dt,
\notag
\\
\bar{L}_j &= \Theta \hat{L}_j \Theta^\dag.
\label{TR-jump}
\end{align}
We note that $\bar{L}_\pm = \hat{L}_\pm$.

\subsection{Detailed fluctuation theorem}

Given a forward QJT $\Gamma = \{ \psi, (t_1, j_1), ..., (t_N, j_N), \varphi \}$,
we construct the corresponding backward trajectory as
$\bar{\Gamma} = \{ \bar{\varphi}, (\bar{t}_1, \bar{\jmath}_1), ..., (\bar{t}_N, \bar{\jmath}_N), \bar{\psi} \}$
with $\ket{\bar{\psi}} = \Theta \ket{\psi}$, $\ket{\bar{\varphi}} = \Theta \ket{\varphi}$,
$\bar{t}_n = \tau - t_{N+1-n}$, and
\begin{align}
\bar{\jmath}_n =
\begin{cases}
(r, l, k) & \text{if}~j_{N+1-n} = (r, k, l)
\\
+ & \text{if}~j_{N+1-n} = -
\\
- & \text{if}~j_{N+1-n} = +.
\end{cases}
\label{jbar}
\end{align}
Then, similarly to Eq.~(\ref{prob_fw_trajectory}),
the probability of $\bar{\Gamma}$ in the time-reversed dynamics is
\begin{align}
\bar{P}_\varepsilon (\bar{\Gamma})
&= \biggl| \bra{\bar{\psi}} \prod_{n=1}^N \Bigl[ \bar{U}_{\text{eff}}(\Delta \bar{t}_n) \bar{L}_{\bar{\jmath}_n} \Bigr] \bar{U}_{\text{eff}}(\Delta \bar{t}_0) \ket{\bar{\varphi}} \biggr|^2 \bar{p}(\bar{\varphi}) dt^N,
\label{prob_bw_trajectory}
\end{align}
where we assume that the spectral decomposition of the ensemble-level initial state in the backward process is given by
$\bar{\rho}_{\text{ini}} = \sum_{\bar{\varphi}} \bar{p}(\bar{\varphi}) \ket{\bar{\varphi}} \bra{\bar{\varphi}}$.

The derivation of the detailed FT is straightforward from Eq.~(\ref{prob_bw_trajectory}).
We substitute Eqs.~(\ref{TR-U_eff}) and (\ref{TR-jump}) and the definitions of the other overlined quantities [shown in Eq.~(\ref{jbar}) and the above] into Eq.~(\ref{prob_bw_trajectory}).
Then, after using $\Theta^\dag \Theta = 1$, we take the complex conjugate of the internal quantity in $|\cdots|^2$.
By using the LDB(-like) conditions (\ref{LDB}) and (\ref{LDB_e}), we finally obtain
\begin{align}
\frac{\bar{P}_\varepsilon (\bar{\Gamma})}{P_\varepsilon (\Gamma)}
&= e^{-\Delta S(\Gamma)} \varepsilon^{N_-(\Gamma) - N_+(\Gamma)},
\label{detailed_FT}
\end{align}
where
\begin{align}
\Delta S(\Gamma) &= \Delta S_{\text{bath}}(\Gamma) + \Delta S_{\text{sys}}(\Gamma),
\label{trajectory_entropy_change}
\\
\Delta S_{\text{bath}}(\Gamma) &= \sideset{}{'}\sum_{n=1}^N \beta_{r_n} \omega_{k_n l_n},
\label{trajectory_bath_entropy_change}
\\
\Delta S_{\text{sys}}(\Gamma) &= -\log \bar{p}(\bar{\varphi}) + \log p(\psi).
\end{align}
Equation (\ref{detailed_FT}) is the detailed FT in our setup.
Here, $\Delta S_{\text{bath}}$ and $\Delta S_{\text{sys}}$ are the trajectory-dependent bath entropy change and system entropy change, respectively.
$\Delta S$ is the total (except for the drain) entropy change.
The sum in Eq.~(\ref{trajectory_bath_entropy_change}) is taken over the bath-induced quantum jumps
by $\hat{L}_{r_n,k_n,l_n}$'s in $\Gamma$ (i.e., the drain-induced jumps are excluded).
$N_-(\Gamma)$ is the number of jumps by $\hat{L}_-$ in $\Gamma$
(which is equal to the number of jumps by $\bar{L}_+$ in $\bar{\Gamma}$),
and $N_+(\Gamma)$ is the number of jumps by $\hat{L}_+$ in $\Gamma$
(which is equal to the number of jumps by $\bar{L}_-$ in $\bar{\Gamma}$).

\subsection{Zero-temperature limit}\label{sec:absolute_irreversibility}

We investigate the zero temperature limit ($\varepsilon \to +0$) at this stage.
To this end, we note that
Eq.~(\ref{prob_fw_trajectory}) leads to $P_\varepsilon(\Gamma) = O(\varepsilon^{N_+(\Gamma)})$
and Eq.~(\ref{prob_bw_trajectory}) leads to $\bar{P}_\varepsilon(\bar{\Gamma}) = O(\varepsilon^{N_-(\Gamma)})$.
Hence, there are two noticeable situations.
One is the case of the QJTs satisfying $N_-(\Gamma) = 0$ and $N_+(\Gamma) > 0$.
In this case, $\lim_{\varepsilon \to +0} P_\varepsilon(\Gamma)$ vanishes while $\lim_{\varepsilon \to +0} \bar{P}_\varepsilon(\bar{\Gamma})$ does not.
The other is the case of $N_-(\Gamma) > 0$ and $N_+(\Gamma) = 0$.
In this case, $\lim_{\varepsilon \to +0} \bar{P}_\varepsilon(\bar{\Gamma})$ vanishes while $\lim_{\varepsilon \to +0} P_\varepsilon(\Gamma)$ does not.
We note that the former case corresponds to the situation in Eq.~(\ref{forward_irreversibility}), and the latter to that in Eq.~(\ref{backward_irreversibility}). 

The trajectories in these situations are called absolutely irreversible \cite{Murashita_etal_14, Ashida_etal_14, Funo_etal_15, Murashita_15, Murashita_Ueda_17, Murashita_etal_17, Hoang_etal_16, Monsel_etal_18, Manikandan_etal_19}.
That is, for the trajectory in the backward (forward) process,
the corresponding forward (backward) trajectory never realizes even stochastically.
The detailed FT (\ref{detailed_FT}) is valid even in these situations in the sense that both sides of the equation diverge or vanish.
To proceed to the integral FT, however, we need careful treatment as explained in the next section.

The origin of the absolute irreversibility in our setup is transparent.
In the limit of $\varepsilon \to +0$, since the photon drain is in the vacuum state,
photons cannot come into the cavity from the drain whereas they can go out of the cavity to the drain.
We see this point in the disappearance of $\mathcal{D}[\hat{L}_+] = \mathcal{D}[\bar{L}_+] = \sqrt{\varepsilon \kappa} \mathcal{D}[\hat{c}^\dag]$ in the QMEs (\ref{L_e}) and (\ref{TR-QME}) in the limit of $\varepsilon \to +0$.
Therefore, if the backward (forward) trajectory contains a jump by $\bar{L}_-$ ($\hat{L}_-$),
it is impossible to generate the corresponding forward (backward) trajectory within the QME.

Note that, if we do not take the zero-temperature limit ($\varepsilon \to +0$),
we can derive the ordinary integral FT from Eq.~(\ref{detailed_FT}): 
\begin{align}
\bigl\langle \exp \bigl[ -\Delta S - (N_- - N_+) E_{\text{ex}}/T_{\text{d}} \bigr] \bigr\rangle = 1.
\label{ordinary_integral_FT}
\end{align}
And, from this FT, we can also derive the ordinary second law:
\begin{align}
\braket{\Delta S} + \braket{N_- - N_+} \frac{E_{\text{ex}}}{T_{\text{d}}} \ge 0.
\label{ordinary_2nd_law}
\end{align}
Here, we used the assignment, $\varepsilon = e^{-E_{\text{ex}} / T_{\text{d}}}$, introduced below Eq.~(\ref{LDB_e}).
However, nonzero but extremely low temperature ($T_{\text{d}} \ll E_{\text{ex}}$) causes the following difficulties.
One is the difficulty in demonstrating the ordinary FT (\ref{ordinary_integral_FT}) in experiments and numerical simulations.
To confirm the validity of the FT (\ref{ordinary_integral_FT}) by averaging over observed trajectories,
it is essential to observe the events (jumps) where the system is excited by heat from the drain since they have extremely large contributions to the FT.
In practice, however, it becomes difficult to observe such events if $E_{\text{ex}} / T_{\text{d}} \gg 1$, because they are extremely rare compared with the opposite events (probability ratio is $e^{-E_{\text{ex}} / T_{\text{d}}}$) due to the LDB-like condition (\ref{LDB_e}).
Another difficulty lies in that the ordinary second law (\ref{ordinary_2nd_law}) is obviously valid but rather useless.
That is, Eq.~(\ref{ordinary_2nd_law}) imposes almost no restriction on $\braket{\Delta S}$, because the second term on the left-hand side is an extremely large positive value if $E_{\text{ex}} / T_{\text{d}} \gg 1$
[note that $\braket{N_- - N_+} > 0$, because $N_- > N_+$ is almost always valid due to the LDB-like condition (\ref{LDB_e}) in this case].
In order to overcome these difficulties arising from the diverging behavior of the drain's contribution,
we have to exclude it and derive an integral FT and a second-law-like inequality that include only $\Delta S$.
We achieve this in the next section.

\section{Integral Fluctuation Theorem}\label{sec:integral_FT}

Here, we derive a modified integral FT in our absolutely irreversible setup.
We use the prescription developed by Murashita \textit{et al.} \cite{Murashita_etal_14, Murashita_15}.
It utilizes some basic ideas in mathematical measure theory \cite{Tao_11}.

\subsection{Decomposition of backward probability measure}

Let $X$ be the set of all QJTs during $[0, \tau]$.
In Table~\ref{tab:subsets}, we summarize the subsets of $X$ which we use in the following argument.
We define the forward probability measure $\mathcal{M}_\varepsilon$ on $X$ and the backward one $\bar{\mathcal{M}}_\varepsilon$ respectively by
\begin{align}
\mathcal{M}_\varepsilon(D\Gamma) &= P_\varepsilon(\Gamma) \mu(D\Gamma),
\\
\bar{\mathcal{M}}_\varepsilon(D\Gamma) &= \bar{P}_\varepsilon(\bar{\Gamma}) \mu(D\Gamma),
\end{align}
where $\mu$ is the Lebesgue measure on $X$.
Then, the ratio in Eq.~(\ref{detailed_FT}) represents the Radon-Nikodym derivative of $\bar{\mathcal{M}}_\varepsilon$ with respect to $\mathcal{M}_\varepsilon$.

\begin{table}[tb]
\caption{\label{tab:subsets}
Definitions of the subsets of $X$ used in the derivation of the modified integral FT.
}
\begin{ruledtabular}
\begin{tabular}{ll}
Definition&
A set of QJTs that have...\\
\colrule
$X_+ = \{ \Gamma \in X \mid N_+(\Gamma) \ge 1 \}$ & At least one jump by $\hat{L}_+$.\\
$X_{0+} = \{ \Gamma \in X \mid N_+(\Gamma) = 0 \}$ & No jump by $\hat{L}_+$.\\
$X_- = \{ \Gamma \in X \mid N_-(\Gamma) \ge 1 \}$ & At least one jump by $\hat{L}_-$.\\
$X_{0-} = \{ \Gamma \in X \mid N_-(\Gamma) = 0 \}$ & No jump by $\hat{L}_-$.\\
$X_{n-} = \{ \Gamma \in X \mid N_-(\Gamma) = n \}$ & Just $n$ jumps by $\hat{L}_-$.
\end{tabular}
\end{ruledtabular}
\end{table}

In the limit of $\varepsilon \to +0$, the absolute irreversibility makes it difficult to define the Radon-Nikodym derivative as seen in the diverging behavior of Eq.~(\ref{detailed_FT}).
In measure theory this difficulty is manifested in the lack of absolute continuity.
We write the limits of the measures as
$\mathcal{M} = \lim_{\varepsilon \to +0} \mathcal{M}_\varepsilon$ and
$\bar{\mathcal{M}} = \lim_{\varepsilon \to +0} \bar{\mathcal{M}}_\varepsilon$.
Let $X_+$ be the set of QJTs that have at least one jump by $\hat{L}_+$ (see Table \ref{tab:subsets}).
As discussed in Sec. \ref{sec:absolute_irreversibility}, for any $\Gamma \in X_+$,
$\lim_{\varepsilon \to +0} P_\varepsilon(\Gamma) = 0$
and $\lim_{\varepsilon \to +0} \bar{P}_\varepsilon(\bar{\Gamma}) > 0$.
Therefore, $\mathcal{M}(X_+) = 0$ and $\bar{\mathcal{M}}(X_+) >0$,
which implies that $\bar{\mathcal{M}}$ is not absolutely continuous with respect to $\mathcal{M}$.

To define the Radon-Nikodym derivative in the absence of absolute continuity,
we must decompose the measure $\bar{\mathcal{M}}$ into the absolutely continuous and singular components.
We can accomplish the decomposition, $\bar{\mathcal{M}} = \bar{\mathcal{M}}^{\text{AC}} + \bar{\mathcal{M}}^{\text{S}}$, with the following restricted measures:
\begin{align}
\bar{\mathcal{M}}^{\text{AC}}(D\Gamma) &= \chi_{0+}(\Gamma) \bar{\mathcal{M}}(D\Gamma),
\label{M_AC}
\\
\bar{\mathcal{M}}^{\text{S}}(D\Gamma) &= \chi_+(\Gamma) \bar{\mathcal{M}}(D\Gamma),
\label{M_S}
\end{align}
where $\chi_\bullet$ ($\bullet = 0+, +$) denotes the indicator function for a subset $X_\bullet$:
\begin{align}
\chi_\bullet(\Gamma) =
\begin{cases}
1 & \Gamma \in X_\bullet
\\
0 & \Gamma \not\in X_\bullet.
\end{cases}
\label{indicator}
\end{align}
$X_+$ is defined in the previous paragraph, and $X_{0+} = \{ \Gamma \in X \mid N_+(\Gamma) = 0 \} = X \backslash X_+$.
Note that $\chi_{0+} = 1 - \chi_+$.
We also note that the Lebesgue decomposition theorem ensures the uniqueness of the decomposition.

We can confirm the absolute continuity of $\bar{\mathcal{M}}^{\text{AC}}$
and the singularity of $\bar{\mathcal{M}}^{\text{S}}$ as follows.
First, from Eq.~(\ref{M_AC}), it is clear that $\bar{\mathcal{M}}^{\text{AC}}(X_+) = 0$ holds.
This implies that $\bar{\mathcal{M}}^{\text{AC}}$ is absolutely continuous with respect to $\mathcal{M}$.
That is, $\bar{\mathcal{M}}^{\text{AC}}(E) = 0$ whenever $\mathcal{M}(E) = 0$.
Next, from Eq.~(\ref{M_S}), it is also clear that $\bar{\mathcal{M}}^{\text{S}}(X_{0+}) = 0$ holds.
This implies that $\bar{\mathcal{M}}^{\text{S}}$ is singular with respect to $\mathcal{M}$.
That is, there exists a set $E \subset X$ such that $\mathcal{M}(E) = 0$ and $\bar{\mathcal{M}}^{\text{S}}(X \backslash E) = 0$ hold.

\subsection{Modified integral FT}

For the absolutely continuous component $\bar{\mathcal{M}}^{\text{AC}}$,
we can define the Radon-Nikodym derivative with respect to $\mathcal{M}$.
Specifically, we have, for any $E \subset X$,
\begin{align}
\bar{\mathcal{M}}^{\text{AC}}(E)
&= \int_E \lim_{\varepsilon \to +0} \chi_{0+}(\Gamma) \bar{\mathcal{M}}_\varepsilon(D\Gamma)
\notag\\
&= \int_E \chi_{0+}(\Gamma) \lim_{\varepsilon \to +0} \frac{\bar{P}_\varepsilon(\bar{\Gamma})}{P_\varepsilon(\Gamma)} \mathcal{M}_\varepsilon(D\Gamma)
\notag\\
&= \int_E \chi_{0+}(\Gamma) \lim_{\varepsilon \to +0} e^{-\Delta S(\Gamma)} \varepsilon^{N_-(\Gamma)}  \mathcal{M}_\varepsilon(D\Gamma)
\notag\\
&= \int_E \chi_{0+}(\Gamma) e^{-\Delta S(\Gamma)} \chi_{0-}(\Gamma) \mathcal{M}(D\Gamma)
\notag\\
&= \int_E e^{-\Delta S(\Gamma)} \chi_{0-}(\Gamma) \mathcal{M}(D\Gamma),
\label{Radon-Nikodym}
\end{align}
where $\chi_\bullet$ ($\bullet = 0+, 0-$) is the indicator function defined in Eq.~(\ref{indicator}).
As defined before, $X_{0+}$ is the set of QJTs that do not include the jumps by $\hat{L}_+$.
On the other hand, $X_{0-}$ is the set of QJTs that do not include the jumps by $\hat{L}_-$ (see Table \ref{tab:subsets}).
In the first line of Eq.~(\ref{Radon-Nikodym}), we used Eq.~(\ref{M_AC}) and the definition of $\bar{\mathcal{M}}$.
In the second line, we used the Radon-Nikodym derivative of $\bar{\mathcal{M}}_\varepsilon$ with respect to $\mathcal{M}_\varepsilon$.
In the third line, we used the detailed FT [Eq.~(\ref{detailed_FT})]
and the fact that $N_+(\Gamma) = 0$ for $\Gamma \in X_{0+}$.
In the fourth line, we used $\lim_{\varepsilon \to +0} \varepsilon^{N_-(\Gamma)} = \chi_{0-}(\Gamma)$.
And in the last line, we used $\chi_{0+}(\Gamma) \mathcal{M}(D\Gamma) = \mathcal{M}(D\Gamma)$, which results from $\mathcal{M}(X_+) = 0$.

Setting $E = X$ in Eq.~(\ref{Radon-Nikodym}),
we obtain
\begin{align}
\bar{\mathcal{M}}^{\text{AC}}(X)
&= \bigl\langle e^{-\Delta S} \bigr\rangle - \Lambda,
\label{Radon-Nikodym_right}
\\
\bigl\langle e^{-\Delta S} \bigr\rangle &= \int_X e^{-\Delta S(\Gamma)} \mathcal{M}(D\Gamma),
\\
\Lambda &= \int_X e^{-\Delta S(\Gamma)} \chi_-(\Gamma) \mathcal{M}(D\Gamma),
\label{big_lambda}
\end{align}
where $\chi_- = 1 - \chi_{0-}$ is the indicator function for $X_- = \{ \Gamma \in X \mid N_-(\Gamma) \ge 1 \} = X \backslash X_{0-}$ (the set of QJTs that have at least one jump by $\hat{L}_-$).

On the other hand,
from $\bar{\mathcal{M}}^{\text{AC}} = \bar{\mathcal{M}} - \bar{\mathcal{M}}^{\text{S}}$
and $\bar{\mathcal{M}}(X) = 1$ (normalization of probability measure),
we have
\begin{align}
\bar{\mathcal{M}}^{\text{AC}}(X) &= 1 - \lambda,
\label{Radon-Nikodym_left}
\\
\lambda &= \int_X \bar{\mathcal{M}}^{\text{S}}(D\Gamma).
\label{small_lambda}
\end{align}
Combining Eqs.~(\ref{Radon-Nikodym_right}) and (\ref{Radon-Nikodym_left}),
we obtain the modified integral FT (\ref{integral_FT}), $\langle e^{-\Delta S} \rangle = 1 - \lambda + \Lambda$, in our setup.
Unlike the ordinary integral FT, this includes the two additional terms,
$\lambda$ and $\Lambda$, on the right-hand side.
As clarified in the next two paragraphs, each of the terms is related to each of the two types of absolute irreversibility mentioned in Introduction and Sec.~\ref{sec:absolute_irreversibility}.
The term of $\lambda$ corresponds to that of the type in Eq.~(\ref{forward_irreversibility})
[the case of $N_-(\Gamma) = 0$ and $N_+(\Gamma) > 0$].
On the other hand, $\Lambda$ corresponds to that of the type in Eq.~(\ref{backward_irreversibility})
[the case of $N_-(\Gamma) > 0$ and $N_+(\Gamma) = 0$].

To see the meaning of $\lambda$, we rewrite Eq.~(\ref{small_lambda}) as
\begin{align}
\lambda = \int_X \chi_+(\Gamma) \bar{\mathcal{M}}(D\Gamma),
\label{small_lambda_another}
\end{align}
where we used the definition (\ref{M_S}) of $\bar{\mathcal{M}}^{\text{S}}$.
This equation implies that $\lambda$ is the probability that
the backward trajectory within the time-reversed QME (\ref{TR-QME}) (in the $\varepsilon \to +0$ limit)
contains at least one jump by $\bar{L}_-$.
In other words, $\lambda$ is the probability that the time-reversed QME (\ref{TR-QME}) generates
the absolutely irreversible backward QJTs of the type in Eq.~(\ref{forward_irreversibility})
(that have no corresponding forward trajectories).

To see the meaning of $\Lambda$, we rewrite Eq.~(\ref{big_lambda}) as
\begin{align}
\Lambda &= \int_X e^{-\Delta S(\Gamma)} \chi_-(\Gamma) \chi_{0+}(\Gamma) \lim_{\varepsilon \to +0} \mathcal{M}_\varepsilon(D\Gamma)
\notag\\
&= \int_X e^{-\Delta S(\Gamma)} \chi_-(\Gamma) \chi_{0+}(\Gamma) \lim_{\varepsilon \to +0} \frac{P_\varepsilon(\Gamma)}{\bar{P}_\varepsilon(\bar{\Gamma})} \bar{\mathcal{M}}_\varepsilon(D\Gamma)
\notag\\
&= \lim_{\varepsilon \to +0} \int_X \chi_-(\Gamma) \chi_{0+}(\Gamma) \varepsilon^{-N_-(\Gamma)} \bar{\mathcal{M}}_\varepsilon(D\Gamma).
\end{align}
In the first line, we used $\chi_{0+}(\Gamma) \mathcal{M}(D\Gamma) = \mathcal{M}(D\Gamma)$ (resulting from $\mathcal{M}(X_+) = 0$) and the definition of $\mathcal{M}$.
In the second line, we used the Radon-Nikodym derivative of $\mathcal{M}_\varepsilon$ with respect to $\bar{\mathcal{M}}_\varepsilon$.
In the last line, we used the reciprocal of Eq.~(\ref{detailed_FT}) with $N_+(\Gamma) = 0$ for $\Gamma \in X_{0+}$.
We can further rewrite this as
\begin{align}
\Lambda &= \lim_{\varepsilon \to +0} \sum_{n \ge 1} \varepsilon^{-n} \Lambda_{\varepsilon, n},
\label{big_lambda_decomposed}
\\
\Lambda_{\varepsilon, n} &= \int_X \chi_{n-}(\Gamma) \chi_{0+}(\Gamma) \bar{\mathcal{M}}_\varepsilon(D\Gamma),
\label{big_lambda_e_n}
\end{align}
where $\chi_{n-}$ is the indicator function for the set $X_{n-}$ of QJTs satisfying $N_-(\Gamma) = n$.
Equation (\ref{big_lambda_e_n}) implies that
$\Lambda_{\varepsilon, n}$ is the probability that
the backward trajectory within the time-reversed QME (\ref{TR-QME}) (before taking the $\varepsilon \to +0$ limit) includes just $n$ jumps by $\bar{L}_+$ and no jump by $\bar{L}_-$.
And $\Lambda$ is the sum of these probabilities weighted by the reciprocals of their order in $\varepsilon$.
In other words, $\Lambda$ is the weighted probability that the time-reversed QME (\ref{TR-QME}) generates the backward QJTs that become absolutely irreversible of the type in Eq.~(\ref{backward_irreversibility}) in the $\varepsilon \to +0$ limit.

We note that both $\lambda$ and $\Lambda$ significantly contribute to the modified integral FT (\ref{integral_FT}). 
This is contrasted to previous works \cite{Murashita_etal_14, Ashida_etal_14, Funo_etal_15, Murashita_15, Murashita_Ueda_17, Murashita_etal_17, Hoang_etal_16, Monsel_etal_18, Manikandan_etal_19}, where only $\lambda$ contributes to their integral FT.
The difference arises from the origins of absolute irreversibility.
As mentioned in Sec.~\ref{sec:absolute_irreversibility}, the origin in our setup lies in asymmetry in the dynamics: cavity photons can go out to the drain but cannot come back. 
In this case, the time-reversed dynamics also induces the irreversibility,
and therefore both $\lambda$ and $\Lambda$ appears.
By contrast, the origin in the previous works lies in restrictions on the forward trajectory.
There, states at some point in the forward trajectories are restricted within a state subspace,
and the subspace is smaller than and contained in the space of sates that can be realized in the backward process.
In this case, only $\lambda$ [the type in Eq.~(\ref{forward_irreversibility})] appears.
In free expansion \cite{Murashita_etal_14}, for example, the initial state is restricted to be a localized state in one side of a box while the backward process allows states in the other side.

\subsection{Second-law-like inequality}

By applying the Jensen inequality, $\braket{e^{-x}} \ge e^{-\braket{x}}$, to the modified integral FT (\ref{integral_FT}),
we obtain the second-law-like inequality in our setup:
\begin{align}
\bigl\langle \Delta S \bigr\rangle \ge -\log (1 -\lambda + \Lambda).
\label{2nd_law}
\end{align}

The right-hand side may be negative,
and as shown in the next subsection this is the case in a certain situation.
We note that $\Delta S$ is the total entropy change without the drain's contribution.
The drain entropy change $\braket{\Delta S_{\text{d}}}$ is always positive infinity
because the drain is in the zero-temperature state
and because it only receives heat from the system (and does not inject heat into the system).
Therefore, the inequality (\ref{2nd_law}) is consistent with the ordinary second law, $\braket{\Delta S + \Delta S_{\text{d}}} \ge 0$, which is the zero-temperature limit ($T_{\text{d}} \to +0$) of Eq.~(\ref{ordinary_2nd_law}).

In our setup, the ordinary second law is valid but rather useless since the left-hand side is infinite.
We may interpret Eq.~(\ref{2nd_law}) as an extraction of a useful finite part from it.
In other words, the second-law-like inequality (\ref{2nd_law}) gives a much better lower bound for $\braket{\Delta S}$ compared with the ordinary second law, $\braket{\Delta S} \ge -\infty$.

\subsection{Estimation of absolute irreversibility}\label{sec:estimation}

Here, we consider a steady-state situation where the initial states in the forward and backward processes are the respective steady states.
That is, we assume $\hat{\rho}_{\text{{ini}}} = \hat{\rho}_{\text{ss}}$ and 
$\bar{\rho}_{\text{{ini}}} = \bar{\rho}_{\text{ss}} = \Theta \hat{\rho}_{\text{ss}} \Theta^\dag$,
where $\hat{\rho}_{\text{ss}}$ is the solution of $\mathcal{L} \hat{\rho}_{\text{ss}} = 0$.
In this situation, we estimate $\lambda$ and $\Lambda$ up to first order in $\kappa \tau$ to find
\begin{align}
\lambda &= \kappa \tau \text{Tr} \bigl[ \hat{c}^\dag \hat{c} \hat{\rho}_{\text{ss}} \bigr] + O[(\kappa \tau)^2],
\label{small_lambda_est}
\\
\Lambda &= \kappa \tau \text{Tr} \bigl[ \hat{c} \hat{c}^\dag \hat{\rho}_{\text{ss}} \bigr] + O[(\kappa \tau)^2].
\label{big_lambda_est}
\end{align}
Thus, $\lambda$ is nearly equal to the average number of photons emitted from the cavity to the drain during $[0, \tau]$.
We give the detailed derivation in Appendix~\ref{sec:derivaton}.

This result leads to $-\lambda + \Lambda \simeq \kappa \tau$.
Therefore, the modified integral FT (\ref{integral_FT}) yields
\begin{align}
\bigl\langle e^{-\Delta S} \bigr\rangle_{\text{ss}} = 1 + \kappa \tau + O[(\kappa \tau)^2],
\label{integral_FT_est}
\end{align}
where the subscript `ss' represents the expectation value for the QJTs in the steady-state situation.
And the second-law-like inequality (\ref{2nd_law}) yields
\begin{align}
\bigl\langle \Delta S_{\text{bath}} \bigr\rangle_{\text{ss}} \ge  -\kappa \tau + O[(\kappa \tau)^2],
\label{2nd_law_est}
\end{align}
where we used $\bigl\langle \Delta S_{\text{sys}} \bigr\rangle_{\text{ss}} = 0$
and $-\log (1 + \kappa \tau + O[(\kappa \tau)^2]) = -\kappa \tau + O[(\kappa \tau)^2]$.
In this condition, $\kappa$ is the only setup parameter on which the modified FT (\ref{integral_FT_est}) and the inequality (\ref{2nd_law_est}) are dependent.

Equation~(\ref{2nd_law_est}) gives a negative lower bound on the average bath entropy change.
But this does not contradict the ordinary second law, as discussed in the previous subsection.

\subsection{Another unravelling}\label{sec:unravelling}

Before closing this section,
we discuss whether the way of stochastic unravelling affects the present results.
As generally known, the QME is invariant to appropriate redefinitions of the jump operators and the Hamiltonian,
and therefore the stochastic unravelling of the QME in terms of QJTs is not unique \cite{Wiseman_Milburn_09}.
In fact, the QME (\ref{QME}) is invariant to the transformation of 
$\kappa \mathcal{D}[\hat{c}] \to \kappa \mathcal{D}[\hat{c} - \alpha]$
[$\hat{L}_- = \sqrt{\kappa} \hat{c} \to \hat{L}_-^\prime = \sqrt{\kappa} (\hat{c} - \alpha)$]
together with $\hat{H} \to \hat{H}' = \hat{H} + (i \kappa / 2) (\alpha^* \hat{c} - \alpha \hat{c}^\dag)$.
Our motivation to discuss this transformation is that
we should use it with $\alpha = \braket{\hat{c}}_{\text{ss}}$
to correctly count the heat to the drain, as we showed previously \cite{Yuge_etal_17}.

To derive the fluctuation theorem for the transformed unravelling,
we replace the additional term $\varepsilon \kappa \mathcal{D}[\hat{c}^\dag]$ in Eq.(\ref{L_e})
with $\varepsilon \kappa \mathcal{D}[\hat{c}^\dag - \alpha^*]$,
which implies $\hat{L}_+ = \sqrt{\varepsilon \kappa} \hat{c}^\dag \to \hat{L}_+^\prime = \sqrt{\varepsilon \kappa} (\hat{c}^\dag - \alpha^*)$.
Then, we obtain a modified integral FT in a similar manner to that in the previous and present sections.
Its form is the same as Eq.~(\ref{integral_FT}) but the transformed jump operators and Hamiltonian are required in $\lambda$ and $\Lambda$.
As a result, the values of $\lambda$ and $\Lambda$ differ from the original ones.

We can clearly see the difference in Eqs.~(\ref{small_lambda_est}) and (\ref{big_lambda_est}).
For the transformed unravelling, we obtain the corresponding results
with the replacement of $\hat{c} \to \hat{c} - \alpha$ and $\hat{c}^\dag \to \hat{c}^\dag - \alpha^*$ in these equations.
Therefore, the values indeed differ from the original ones at first order in $\kappa \tau$.

As easily shown, however, $-\lambda + \Lambda \simeq \kappa \tau$ remains valid up to first order.
That is, in this condition, the modified integral FT (\ref{integral_FT_est}) and second-law-like inequality (\ref{2nd_law_est}) are invariant to the transformation.

\section{Example}\label{sec:example}

In this section, we numerically demonstrate our results in a model of cQED system.
Specifically, we investigate the steady-state situation and show that the modified integral FT and the second-law-like inequality in the small $\kappa \tau$ region, Eqs.~(\ref{integral_FT_est}) and (\ref{2nd_law_est}), are valid in the model.

\begin{figure}[t]
\begin{center}
\includegraphics[width=\linewidth]{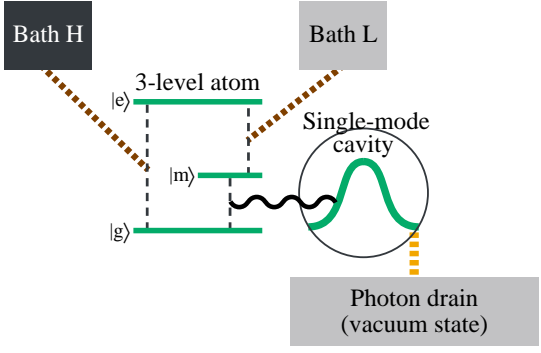}
\end{center}
\caption{Schematic draw of the example model.
The cavity QED system is composed of a three-level atom and a single-mode cavity.
}
\label{fig:model}
\end{figure}

\subsection{Model}

The model is a cavity QED version of a quantum heat engine proposed by Scovil and Shultz-DuBois \cite{Scovil_Schulz-DuBois_59}.
We show a schematic setup of the model in Fig.~\ref{fig:model}.

In this model, the cQED system is composed of a three-level atom and a single-mode cavity.
The lowest, middle, and highest levels of the atom are denoted by $\ket{\text{g}}$, $\ket{\text{m}}$, and $\ket{\text{e}}$, respectively.
The cavity mode interacts with the atom through the transition between $\ket{\text{g}}$ and $\ket{\text{m}}$, and the cavity frequency is resonant to this transition frequency $\omega_{\text{m}}$.
The Hamiltonian of the cQED system is a variant of the Jaynes-Cummings Hamiltonian:
\begin{align}
\hat{H} = \sum_{q=\text{m,e}} \omega_q \ket{q}\bra{q} + \omega_{\text{m}} \hat{c}^\dag \hat{c}
+ g_{\text{c}} \Bigl( \ket{\text{m}}\bra{\text{g}} \hat{c} + \ket{\text{g}}\bra{\text{m}} \hat{c}^\dag \Bigr),
\label{model_hamiltonian}
\end{align}
where $\omega_{\text{e}}$ is the transition frequency between $\ket{\text{g}}$ and $\ket{\text{e}}$,
and $g_{\text{c}}$ is the coupling constant.
We can exactly determine the eigenstates of this Hamiltonian:
\begin{align*}
\ket{E_k} =
\begin{cases}
\ket{0,\text{g}} & k=(0,\text{g})\\
\bigl( \ket{n-1,\text{m}} \pm \ket{n,\text{g}} \bigr)/\sqrt{2} & k=(n,\pm),~n \ge 1\\
\ket{n,\text{e}} & k=(n,\text{e}),~~n \ge 0.
\end{cases}
\end{align*}
Here, $\ket{n,q} = \ket{n}\ket{q}$ represents that the cavity is in the $n$-photon number state $\ket{n}$ and the atom is in the state $\ket{q}$ ($q=\text{g, m, e}$).
The corresponding energy eigenvalues are
\begin{align*}
E_k =
\begin{cases}
0 & k=(0,\text{g})\\
n \omega_{\text{m}} \pm g_{\text{c}} \sqrt{n} & k=(n,\pm),~n \ge 1\\
n \omega_{\text{m}} + \omega_{\text{e}} & k=(n,\text{e}),~~n \ge 0.
\end{cases}
\end{align*}

The environment is composed of two heat baths and a photon drain.
The higher-temperature bath (denoted by H) is thermally coupled to the atom 
through the $\ket{\text{g}} \leftrightarrow \ket{\text{e}}$ transition,
and the lower-temperature bath (denoted by L) is thermally coupled to the atom 
through the $\ket{\text{m}} \leftrightarrow \ket{\text{e}}$ transition.
The photon drain is as described in the general setup (Sec.~\ref{sec:setup}).

Using the Born-Markov approximation and the rotating-wave approximation and neglecting the Lamb shift term \cite{Breuer_Petruccione_02}, we can derive the QME for the cQED system subject to this environment.
The Liouvillian $\mathcal{L}$ of the QME is in the form of Eq.~(\ref{QME}) with
$\hat{H}$ given by Eq.~(\ref{model_hamiltonian}) and the bath-induced jump operators given by
\begin{align*}
\hat{L}_{\text{H},k,l} &=
\begin{cases}
\sqrt{\nu_{\text{H}}^-(\omega_{kl})}
\ket{E_l} \braket{E_l | \text{g}}\braket{\text{e} | E_k} \bra{E_k}
& \omega_{kl} > 0
\\
\sqrt{\nu_{\text{H}}^+(\omega_{kl})}
\ket{E_l} \braket{E_l | \text{e}} \braket{\text{g} | E_k} \bra{E_k}
& \omega_{kl} < 0,
\end{cases}
\\
\hat{L}_{\text{L},k,l} &=
\begin{cases}
\sqrt{\nu_{\text{L}}^-(\omega_{kl})}
\ket{E_l} \braket{E_l | \text{m}}\braket{\text{e} | E_k} \bra{E_k}
& \omega_{kl} > 0
\\
\sqrt{\nu_{\text{L}}^+(\omega_{kl})}
\ket{E_l} \braket{E_l | \text{e}} \braket{\text{m} | E_k} \bra{E_k}
& \omega_{kl} < 0.
\end{cases}
\end{align*}
Here, $\nu_r^-(\omega) = \gamma_r |\omega|^3 [N_r(\omega) + 1]$
and $\nu_r^+(\omega) = \gamma_r |\omega|^3 N_r(-\omega)$.
$N_r(\omega) = 1/[\exp(\beta_r \omega) - 1]$ is the Bose distribution with the inverse temperature $\beta_r$,
and $\gamma_r |\omega_{kl}|^3$ represents the spontaneous transition rate between $\ket{E_k}$ and $\ket{E_l}$
due to the coupling to Bath $r$ ($r=\text{H, L}$).
As easily shown, these jump operators satisfy the LDB condition (\ref{LDB}).

\subsection{Numerical result}

\begin{figure}[t]
\begin{center}
\includegraphics[width=\linewidth]{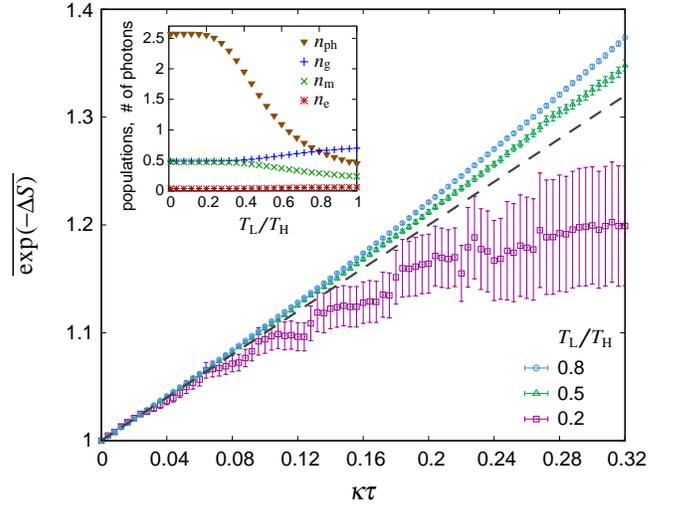}
\end{center}
\caption{Final-time ($\tau$) dependence of the average $\overline{e^{-\Delta S}}$ over QJTs for $T_{\text{L}} / T_{\text{H}} = 0.8$ (circles), $0.5$ (triangles), and $0.2$ (squares).
The error bars show $\pm$SE.
The dashed line represents $1 + \kappa \tau$ [the right-hand side of Eq.~(\ref{integral_FT_est})].
In the units of $\omega_{\text{m}} = 1$, the parameters are set to be $\omega_{\text{e}} = 2.4$, $g_{\text{c}}=0.001$, $\kappa = 0.0004$, $\gamma_{\text{H}} = \gamma_{\text{L}} = 0.01$, and $T_{\text{H}} = 1$.
The cutoff number of cavity photons is set to be 10.
The sample number of QJTs is $1.5 \times 10^6$ for each $T_{\text{L}} / T_{\text{H}}$.
The inset shows the steady-state expectation values of the photon number $n_{\text{ph}}$ and the level populations $n_q$ ($q = \text{g, m, e}$), plotted against $T_{\text{L}} / T_{\text{H}}$.
}
\label{fig:result_1}
\end{figure}

We first investigate basic properties of the model in the steady state.
To this end, we numerically solve the steady-state equation $\mathcal{L} \hat{\rho}_{\text{ss}} = 0$ and calculate the expectation values, $n_{\text{ph}} = \text{Tr}[\hat{\rho}_{\text{ss}} \hat{c}^\dag \hat{c}]$ and $n_q = \text{Tr}[\hat{\rho}_{\text{ss}} \ket{q} \bra{q}]$ ($q = \text{g, m, e}$). 
In the inset of Fig.~\ref{fig:result_1}, we plot these expectation values as functions of the temperature $T_{\text{L}} = 1 / \beta_{\text{L}}$ of Bath L (with the temperature $T_{\text{H}} = 1 / \beta_{\text{H}}$ of Bath H fixed).
As lowering $T_{\text{L}}$, 
we observe that $n_{\text{ph}}$ becomes larger than one together with $n_{\text{g}}$ and $n_{\text{m}}$ approaching to each other.
This behavior indicates that the system crosses over from a nearly thermal state (at $T_{\text{L}} \simeq T_{\text{H}}$) to a lasing state (at $T_{\text{L}} \ll T_{\text{H}}$).

Next, we test the validity of the modified integral FT (\ref{integral_FT_est}) and the second-law-like inequality (\ref{2nd_law_est}) in the steady-state situation and in the region of small $\kappa \tau$.
For this purpose, we perform the Monte Carlo simulation with the stochastic wave function method of generating many QJTs \cite{Breuer_Petruccione_02, Carmichael_93, Wiseman_Milburn_09}.
We calculate the entropy change $\Delta S$ in Eq.~(\ref{trajectory_entropy_change}) for each QJT
and take the averages, $\overline{e^{-\Delta S}}$ and $\overline{\Delta S}$, over the QJTs.
In the simulation, we make two assumptions to describe the steady-state situation in Sec.~\ref{sec:estimation}.
One is that the ensemble-level initial state is the steady state $\hat{\rho}_{\text{ss}} = \sum_\psi p_{\text{ss}}(\psi) \ket{\psi} \bra{\psi}$, so that we choose one of $\{ \ket{\psi} \}_\psi$ as the initial wave function according to the probability $p_{\text{ss}}$.
The other is that the basis set of the measurement at the final time $\tau$ is also $\{ \ket{\psi} \}_\psi$.

In the main panel of Fig.~\ref{fig:result_1}, we show the numerical results of $\overline{e^{-\Delta S}}$ against $\kappa \tau$ for three cases of $T_{\text{L}}$ chosen from the nearly thermal regime ($T_{\text{L}} / T_{\text{H}} = 0.8$; circles), the crossover regime ($T_{\text{L}} / T_{\text{H}} = 0.5$; triangles), and the lasing regime ($T_{\text{L}} / T_{\text{H}} = 0.2$; squares).
We also plot $1 + \kappa \tau$ [the right-hand side of Eq.~(\ref{integral_FT_est})] as the dashed line.
In all the cases, $1 + \kappa \tau$ is within the standard error (SE) of $\overline{e^{-\Delta S}}$
in the small $\kappa \tau$ region ($\kappa \tau \lesssim 0.04$).
Therefore, we numerically confirm the validity of the integral FT (\ref{integral_FT_est}) to first order in $\kappa \tau$.

\begin{figure}[t]
\begin{center}
\includegraphics[width=\linewidth]{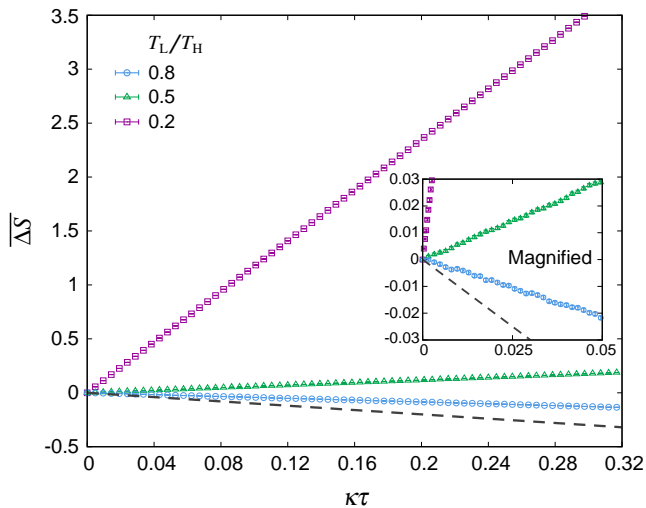}
\end{center}
\caption{Final-time ($\tau$) dependence of the average $\overline{\Delta S}$ over QJTs for $T_{\text{L}} / T_{\text{H}} = 0.8$ (circles), $0.5$ (triangles), and $0.2$ (squares).
The error bars show $\pm$SE (most of which are smaller than the symbols).
The dashed line represents $-\kappa \tau$ [the right-hand side of Eq.(\ref{2nd_law_est})].
The inset is a magnified plot for $\kappa \tau < 0.05$.
The parameter values are the sames as those in Fig.~\ref{fig:result_1}.
}
\label{fig:result_2}
\end{figure}

In Fig.~\ref{fig:result_2}, we show the results of $\overline{\Delta S}$ against $\kappa \tau$ for the three cases of $T_{\text{L}}$.
We also plot $-\kappa \tau$ [the right-hand side of Eq.(\ref{2nd_law_est})] as the dashed line.
In all the cases, $\overline{\Delta S}$ is larger than $-\kappa \tau$.
For $T_{\text{L}} / T_{\text{H}} = 0.8$, in particular, though $\overline{\Delta S}$ is negative, it exceeds $-\kappa \tau$ (see the magnified plot in the inset of Fig.~\ref{fig:result_2}).
For larger $T_{\text{L}} / T_{\text{H}}$ (even for  $T_{\text{L}} / T_{\text{H}} \ge 1$), we also observe that the numerical results satisfy $\overline{\Delta S} \ge -\kappa \tau$ (not shown in the figure).
Therefore, we numerically confirm the validity of the second-law-like inequality (\ref{2nd_law_est}) to first order in $\kappa \tau$.

\section{Conclusion}\label{sec:conclusion}

In the present paper, we have analyzed a general setup of open cQED systems
to investigate the FT for a case where a heat bath (called photon drain) is in a zero-temperature state.
We have found that the diverging behavior in entropy change---caused by the zero-temperature property of the drain---is manifestations of absolute irreversibility.
Using the method for the absolutely irreversible situations,
we have derived the modified integral FT (\ref{integral_FT}) and the second-law-like inequality (\ref{2nd_law}).
Two modification terms appear in these results---one (denoted by $\lambda$) is known in previous works \cite{Murashita_etal_14, Ashida_etal_14, Funo_etal_15, Murashita_15, Murashita_Ueda_17, Murashita_etal_17, Hoang_etal_16, Monsel_etal_18, Manikandan_etal_19}, whereas the other (denoted by $\Lambda$) is discovered in the present work for the first time.
$\lambda$ is the probability of the backward trajectories in the time-reversed dynamics whose corresponding forward trajectories cannot be generated within the forward dynamics.
$\Lambda$ is related to the probability of the forward trajectories in the forward dynamics whose corresponding backward trajectories cannot be generated within the reversed dynamics.

We have also shown that, in a steady-state and small cavity-loss condition, the modification terms are directly connected to the average number of photons emitted to the drain.
Using this result, we have estimated the modified integral FT and the second-law-like inequality to linear order in $\kappa \tau$.
Their approximate forms include only $\kappa \tau$ and are thus independent of the setup details.
Moreover, we have shown that these forms in this condition are invariant to the change of stochastic unravelling though it remains open whether this is the case in other conditions.

We have illustrated the results in a model of quantum heat engine.
Using the stochastic wave function method,
we have numerically confirmed the validity of the modified FT (\ref{integral_FT_est}) and the inequality (\ref{2nd_law_est}).

The present results give several suggestions on quantum thermodynamics in open cQED systems.
We should remove the drain's contribution from the entropy change in order to avoid difficulty of diverging behavior and to obtain meaningful results of the FT and second-law-like inequality.
At the expense of the removal, we should take the modification terms into consideration.
The second-law-like inequality, in particular, indicates that the average of the entropy change (without the drain's contribution) may be negative.
The numerical result in the example suggests that the negative entropy change is more likely in nearly thermal situations.

\begin{acknowledgments}
The authors acknowledge Yasuhiro Yamada for helpful discussion.
This work was supported by JSPS KAKENHI Grant Number JP18K03454.
\end{acknowledgments}

\appendix

\section{Derivation of Eqs.~(\ref{small_lambda_est}) and (\ref{big_lambda_est})}
\label{sec:derivaton}

In this appendix, we mainly give the detailed derivation for the estimation (\ref{small_lambda_est}) of $\lambda$.
We can derive the estimation (\ref{big_lambda_est}) of $\Lambda$ in a similar manner.

As explained below Eq.~(\ref{small_lambda_another}),
$\lambda$ is the probability that the jump by $\bar{L}_-$ occurs at least once in the backward trajectory within the time-reversed QME (\ref{TR-QME}) (in the $\varepsilon \to +0$ limit).
Therefore, we can decompose $\lambda$ as $\lambda = \sum_{n \ge 1} \lambda_n$, where $\lambda_n$ is the probability that the jump by $\bar{L}_-$ occurs $n$ times.
We note that $\lambda_n = O[(\kappa \tau)^n]$ because each single jump by $\bar{L}_-$ induces $\kappa$ in the probability.
Therefore, in order to estimate $\lambda$ to first order in $\kappa \tau$, it is sufficient to investigate $\lambda_1$.

From Eq.~(\ref{prob_bw_trajectory}), we can write $\lambda_1$ as
\begin{widetext}
\begin{align}
\lambda_1
&= \text{Tr} \sum_{N=0}^\infty \sideset{}{'}\sum_{j_1, ..., j_N} 
\sum_{N'=0}^\infty \sideset{}{'}\sum_{j'_1, ..., j'_{N'}}
\int_0^\tau \int_0^{t_N} \dotsi \int_0^{t_2} \int_0^{t_1}
\int_0^{t_-} \int_0^{t'_{N'}} \dotsi \int_0^{t'_2}
\notag\\
& \times
\prod_{n=1}^N \Bigl[ \bar{\mathcal{U}}_{\text{eff}}(\Delta t_n) \bar{\mathcal{J}}_{j_n} \Bigr] \bar{\mathcal{U}}_{\text{eff}}(\Delta t_0) 
\bar{\mathcal{J}}_-
\prod_{n=1}^N \Bigl[ \bar{\mathcal{U}}_{\text{eff}}(\Delta t'_n) \bar{\mathcal{J}}_{j_n} \Bigr] \bar{\mathcal{U}}_{\text{eff}}(\Delta t'_0) 
\bar{\rho}_{\text{ini}}.
\label{lambda_1}
\end{align}
\end{widetext}
Here, we abbreviate the overlines on $j$'s and $t$'s for notational simplicity.
The primed summation symbols stand for sums over bath-induced quantum jumps
[i.e., $j_n^{(\prime)}$ is one of $(r,k,l)$'s].
And $\Delta t_n^{(\prime)} = t_{n+1}^{(\prime)} - t_n^{(\prime)}$, where $t_{N+1} = \tau$, $t_0 = t'_{N+1} = t_-$ (time at which the jump by $\bar{L}_-$ occurs), and $t'_0 = 0$.
Note that we take the $\varepsilon \to +0$ limit for $\bar{\mathcal{U}}_{\text{eff}}$ in Eq.~(\ref{lambda_1}) [though we defined it in Eq.~(\ref{TR-U_eff}) for the case of $\varepsilon > 0$].

We now consider the steady-state situation. Then, we can set the initial state of the backward trajectory as $\bar{\rho}_{\text{ini}} = \bar{\rho}_{\text{ss}} = \Theta \hat{\rho}_{\text{ss}} \Theta^\dag$.
Moreover, we note that $\bar{\mathcal{J}}_-$ in Eq.~(\ref{lambda_1}) contains $\kappa$.
Hence, it is sufficient to estimate the remaining part of the equation to zeroth order in $\kappa$.
That is, in Eq.~(\ref{lambda_1}) we can approximately replace $\bar{\mathcal{U}}_{\text{eff}}$ and $\bar{\rho}_{\text{ini}} = \bar{\rho}_{\text{ss}}$ with the zeroth-order ones, which are denoted by $\bar{\mathcal{U}}_{\text{eff}}^0$ and $\bar{\rho}_{\text{ss}}^0$, respectively.
Then, by using an expression similar to Eq.~(\ref{time_ev_ens}), we can rewrite the equation as
\begin{align}
\lambda_1 = \int_0^\tau \text{Tr} \Bigl[ e^{\bar{\mathcal{L}}^0 (\tau - t_-)} \bar{\mathcal{J}}_-
e^{\bar{\mathcal{L}}^0 t_-} \bar{\rho}_{\text{ss}}^0 \Bigr] + O[(\kappa \tau)^2],
\end{align}
where $\bar{\mathcal{L}}^0$ is the Liouvillian of the time-reversed QME for the case of $\kappa = 0$
[i.e., Eq.~(\ref{TR-QME}) with $\kappa = 0$].
Furthermore, by noting $\bar{\mathcal{L}}^0 \bar{\rho}_{\text{ss}}^0 = 0$ and the trace-preserving property of the QME, we obtain
\begin{align}
\lambda_1 &= \int_0^\tau \text{Tr} \bigl[ \bar{\mathcal{J}}_- \bar{\rho}_{\text{ss}}^0 \bigr] + O[(\kappa \tau)^2]
\notag\\
&= \kappa \int_0^\tau \text{Tr} \bigl[ \hat{c} \bar{\rho}_{\text{ss}}^0 \hat{c}^\dag \bigr] dt_- + O[(\kappa \tau)^2]
\notag\\
&= \kappa \tau \text{Tr} \bigl[ \hat{c}^\dag \hat{c} \hat{\rho}_{\text{ss}}^0 \bigr] + O[(\kappa \tau)^2],
\end{align}
where we used $\bar{\rho}_{\text{ss}}^0 = \Theta \hat{\rho}_{\text{ss}}^0 \Theta^\dag$ and the cyclic property of the trace.
In this equation we can replace $\hat{\rho}_{\text{ss}}^0$ with $\hat{\rho}_{\text{ss}}$ to first order in $\kappa \tau$,
so that we finally obtain Eq.~(\ref{small_lambda_est}).

The derivation of Eq.~(\ref{big_lambda_est}) is similar to that of Eq.~(\ref{small_lambda_est}).
In order to estimate $\Lambda$ to first order in $\kappa \tau$,
it is sufficient to calculate $\lim_{\varepsilon \to +0} \varepsilon^{-1} \Lambda_{\varepsilon,1}$ in Eq.~(\ref{big_lambda_decomposed}).
We can estimate $\Lambda_{\varepsilon,1}$ in a parallel way to that of $\lambda_1$ by simply replacing $\bar{\mathcal{J}}_-$ with $\bar{\mathcal{J}}_+$, so that we obtain
\begin{align}
\Lambda_{1,\varepsilon} = \varepsilon \kappa \tau \text{Tr} \bigl[ \hat{c} \hat{c}^\dag \hat{\rho}_{\text{ss}}^0 \bigr] + O[(\kappa \tau)^2].
\end{align}
We thus derive Eq.~(\ref{big_lambda_est}).

\end{document}